# Fluctuation – dissipation physics of stimulated light scattering from laser-driven density gratings

Aleksei Zheltikov

Institute for Quantum Science and Engineering, Department of Physics and Astronomy, Texas A&M University, College Station, Texas 77843; zheltikov@tamu.edu

The fluctuation – dissipation theorem (FDT) is shown to provide a powerful resource for the analysis of stimulated light scattering from laser-driven density gratings, including stimulated Brillouin scattering (SBS) and its kinetic-regime extension. In the physical setting of stimulated light scattering by density gratings, the FDT establishes that the dynamics of disturbances induced in a medium by a laser-driven electrostrictive force unfolds via the same physical pathways as the dynamics of internal, spontaneous fluctuations in this medium at equilibrium. When integrated into a suitable kinetic framework, the FDT leads to a significant simplification of the analysis of stimulated light scattering, allowing the stimulated gain/loss spectrum to be found directly from the spectrum of spontaneous density fluctuations without the need to solve kinetic equations with an external-field term. Such FDT-based framework enables an accurate description of stimulated light scattering within a vast parameter space, recovering all the signature properties of sound-wave-mediated Stokes amplification in the hydrodynamic regime, providing a continuous crossover from the hydrodynamic to kinetic regime of stimulated scattering, and explaining distinctly different properties of kinetic stimulated scattering from density gratings, consistent with experiments on stimulated scattering in moderate-pressure gases. As important physical benchmarks, in the limits of vanishingly low and very high collision frequencies, this solution for the SBS gain recovers the FDT-transformed solutions of, respectively, the Vlasov and Navier – Stokes equations. This treatment sheds light on how the causality of density fluctuations is manifested in stimulated scattering and how the universality of Doppler broadening in the spectrum of spontaneous density fluctuations in the kinetic regime translates into the spectra of stimulated and coherent scattering.

## 1. Introduction

In a standard, widely accepted nomenclature of light – matter interactions [1 – 4], the processes of light scattering by low-frequency material excitations are isolated from Raman scattering and are treated separately, with suitable physical models and adequate theoretical formalisms [2 – 8]. Of special prominence among such processes are Brillouin scattering [2 – 4, 6], broadly defined as light scattering by acoustic waves, or isoentropic adiabatic density fluctuations, Rayleigh scattering [3, 4, 9], understood in the most general sense as scattering by entropy, or isobaric density fluctuations, and Rayleigh-wing scattering [3, 4], caused by molecular orientation fluctuations. In high-intensity laser fields, all these processes can be observed in stimulated and coherent regimes, providing a drastic enhancement of the scattering process.

Since the discovery of stimulated Brillouin scattering (SBS) at the dawn of the laser era [10], the nonlinear regimes of Brillouin and Rayleigh scattering have found a vast variety of applications, spanning from nonlinear spectroscopy [2, 11 – 14], fiber-optic sensing [15 – 18], bioimaging [19 – 23], tunable all-optical delay and slow-light generation [24 – 26], and diagnostics of high-speed flows and fluid dynamics [27 – 36]. Alongside its Raman counterpart, stimulated Brillouin scattering is among the most important parametric instability processes in laser-produced plasmas [37 – 41], including plasmas of laser-driven inertial confinement fusion [42 – 45]. In a suitably engineered laser–plasma interaction setting, on the other hand, SBS can provide a mechanism for an efficient pulse amplification and compression [46 – 54], opening routes toward sources of high-intensity subfemtosecond x-ray pulses [55]. Prominent applications of SBS in advanced laser technologies include creation of phase-conjugation mirrors [56 – 59], narrow-linewidth lasers [60 – 65], as well as compressors, beam combiners, and amplifiers of high-energy laser pulses [66 – 71], including gas-phase pulse compressors and beam combiners for the latest generation of high-energy laser facilities for inertial confinement fusion [72 – 77].

In solids, liquids, and high-pressure gases, low-frequency light scattering is adequately understood [1 – 5, 11] in terms of scattering by dielectric permittivity variations – random in the spontaneous scattering or regular, via laser-induced density gratings, in coherent and stimulated regimes. A standard framework for the analysis of such a process in solids and fluids is based on a set of hydrodynamic equations consisting of the Navier – Stokes equation with an electrostrictive force, continuity equation, and the heat-conduction equation with a source term including absorption of laser radiation [2 – 4, 11]. In moderate-pressure gases, however, where the mean free path of gas particles is longer than the effective wavelength of the density wave transferring the energy of the laser driver to the scattered field, the entire picture of light scattering breaks, as indicated by drastic changes in the properties of scattered light and the shape of the gain spectrum [6, 7, 27 – 30]. In this regime, light scattering can no longer be viewed as scattering by continuous fluid but should be treated instead as scattering by an ensemble of independent, free-streaming particles. This regime of scattering, often referred to as the kinetic or dilute-gas regime, has long been a subject of much interest in the context of applications of stimulated and coherent scattering in nonlinear spectroscopy of atomic and molecular gases [6, 7, 27 – 30, 78 – 81], laser-based gas-flow diagnostics [31 – 34, 82, 83], and, most recently, pulse compression in laser drivers for inertial fusion [76, 77, 84].

As the gas ceases to be a continuous fluid in a low- or moderate-pressure regime, it can no longer be described hydrodynamically but should be treated using suitably adapted kinetic equations. Specifically, the properties of spontaneous low-frequency light scattering are adequately understood [85 – 91] in terms of the spectrum of density fluctuations found by solving the Boltzmann equation with a suitable linearization of the collision operator, e.g., using the single-relaxation-time linear Bhatnagar – Gross – Krook (BGK) approximation [92] or models with multiple independent relaxation rates, such as the Shakhov model [93, 94] or the Tenti models [95 – 97]. Extension of this framework to stimulated and coherent regimes of light – matter interaction has been achieved by adding field-dependent source terms to the linearized Boltzmann equation (LBE) in one of these approximations.

Such generalizations have been shown to provide an accurate description of experiments on coherent Rayleigh [80] and coherent Rayleigh – Brillouin scattering [32 – 34], as well as experiments on stimulated low-frequency light scattering [84]. However, because these solutions rederive the spectrum of density fluctuations from the field-modified LBE, expressing this spectrum as the ratio of complex polynomials of the plasma dispersion function, they do not always provide a clear, physically intuitive insight into how light-driven density fluctuations give rise to light-intensity-dependent gain and coherence buildup in stimulated and coherent regimes of light scattering. Among the fundamental physical questions that remain to be addressed are the questions of how the causality of density fluctuations is manifested in the spectra of stimulated scattering and how the universality of Doppler-broadened spectra emerging in spontaneous scattering off density fluctuations translates into the spectra of stimulated and coherent scattering. Standing out among unanswered practical questions is the question of whether the kinetic regime of stimulated scattering from laser-driven density gratings can provide any practically significant gain enhancement for efficient pulse compression in high-power laser beamlines, including laser systems for inertial fusion [72 – 77].

Here, seeking to address these questions, we resort to the fluctuation – dissipation theorem (FDT). We show that, when integrated into a suitable kinetic framework, FDT-based treatment enables an accurate description of stimulated light scattering within a vast parameter space, recovering all the signature properties of sound-wave-mediated Stokes amplification in the hydrodynamic regime, providing a continuous crossover from the hydrodynamic to kinetic regime of stimulated scattering, and explaining distinctly different properties of kinetic stimulated scattering from density gratings, consistent with experiments on stimulated scattering in moderate-pressure gases.

## 2. Electrostriction-induced polarization and stimulated gain

We consider a standard SRBS/CRBS setting [2, 3] in which a laser field,

$$\mathbf{E}(\mathbf{r},\mathrm{t}) = \frac{1}{2}\mathbf{E}_p(\mathbf{r},\mathrm{t})\exp(i\phi_p) + \frac{1}{2}\mathbf{E}_s(\mathbf{r},\mathrm{t})\exp(i\phi_s) + c.c., \quad (1)$$

consisting of pump and Stokes/seed laser fields with field amplitudes $\mathbf{E}_p(\mathbf{r},\mathrm{t})$ and $\mathbf{E}_s(\mathbf{r},\mathrm{t})$, phases $\phi_p = \mathbf{k}_p \cdot \mathbf{r} - \omega_p t$ and $\phi_s = \mathbf{k}_s \cdot \mathbf{r} - \omega_s t$, wave vectors $\mathbf{k}_p$ and $\mathbf{k}_s$, and central frequencies $\omega_p$ and $\omega_s$, induces a ponderomotive potential

$$U_\mathrm{p}(\mathbf{r},\mathrm{t}) = -\frac{\alpha}{2}[\mathbf{E}(\mathbf{r},\mathrm{t})]^2, \quad (2)$$

where $\alpha$ is the polarizability of gas particles.

This potential gives rise to a ponderomotive force

$$\mathbf{F}(\mathbf{r},\mathrm{t}) = -\boldsymbol{\nabla} U_\mathrm{p}(\mathbf{r},\mathrm{t}) = \frac{\alpha}{2}\boldsymbol{\nabla}[\mathbf{E}(\mathbf{r},\mathrm{t})]^2, \quad (3)$$

which pushes gas particles toward regions of higher field strength, thus inducing a change $\delta\rho(\mathbf{r},t)$ in gas density $\rho(\mathbf{r},t)$. The resulting change $\delta\varepsilon(\mathbf{r},t)$ in dielectric permittivity $\varepsilon(\mathbf{r},t)$ couples the pump and Stokes fields via the nonlinear polarization, which, in the first order in $\delta\rho(\mathbf{r},t)$, i.e., with

$$\delta\varepsilon(\mathbf{r},t) \approx \left(\frac{\partial\varepsilon}{\partial\rho}\right)\delta\rho(\mathbf{r},t), \quad (4)$$

is given by [2, 3]

$$\mathbf{P}^{(3)}(\mathbf{r},\mathrm{t}) = \frac{\delta\varepsilon(\mathbf{r},t)}{4\pi}\mathbf{E}(\mathbf{r},t) = \frac{1}{4\pi}\left(\frac{\partial\varepsilon}{\partial\rho}\right)\delta\rho(\mathbf{r},t)\mathbf{E}(\mathbf{r},t). \quad (5)$$

When treated in the first order in $\delta\rho$, this nonlinear polarization is cubic in the laser field $\mathbf{E}(\mathbf{r},t)$ because $\delta\rho$ is quadratic in $\mathbf{E}(\mathbf{r},t)$.

Nonlinear coupling between the pump and Stokes fields and nonlinear-polarization wave opens a channel for energy transfer from the pump to the Stokes field, as described by the standard coupled-wave equations for the slowly varying amplitudes $\mathbf{E}_p(\mathbf{r},\mathrm{t})$, $\mathbf{E}_s(\mathbf{r},\mathrm{t})$, and $\delta\rho(\mathbf{r},t)$ [2, 3, 5]. Searching for $\delta\rho(\mathbf{r},t)$ and $\mathbf{P}^{(3)}(\mathbf{r},\mathrm{t})$ in the form of $\delta\rho(\mathbf{r},t) = \delta\rho_{\mathbf{K},\Omega}\exp(i\Phi) + \delta\rho^*_{\mathbf{K},\Omega}\exp(-i\Phi)$ and $\mathbf{P}^{(3)}(\mathbf{r},t) = \mathbf{P}^{(3)}_{\mathbf{K},\Omega}\exp(i\Phi) + \mathbf{P}^{(3)*}_{\mathbf{K},\Omega}\exp(-i\Phi)$, with $\mathbf{K} = \mathbf{k}_p - \mathbf{k}_s$ and $\Omega = \omega_p - \omega_s$, and $\Phi = \mathbf{K}\cdot\mathbf{r} - \Omega t$, we find from Eq. (5)

$$\mathbf{P}^{(3)}_{\mathbf{K},\Omega} = \frac{\gamma_e}{4\pi\rho_0}\delta\rho^*_{\mathbf{K},\Omega}\mathbf{E}_p = \chi^{(3)}(\mathbf{K},\Omega)\mathbf{E}_p\left(\mathbf{E}^*_{\mathrm{p}}\cdot\mathbf{E}_s\right), \quad (6)$$

where $\gamma_e = \rho_0\left(\frac{\partial\varepsilon}{\partial\rho}\right) = \varepsilon - 1 = 4\pi\mathcal{N}\alpha$ is the electrostrictive coefficient, $\mathcal{N}$ is the number density of gas particles, and $\chi^{(3)}(\mathbf{K},\Omega)$ is the third-order nonlinear susceptibility. Because, by the very nature of this process, $\Omega \ll \omega_p, \omega_s$ and $k_p = |\mathbf{k}_p| \approx k_s = |\mathbf{k}_s|$, $K = |\mathbf{K}| \approx 2k_p\sin(\theta/2)$, where $\theta$ is the angle between $\mathbf{k}_p$ and $\mathbf{k}_s$. Specifically, for counterpropagating pump and Stokes beams, $\theta = \pi$, we find $K \approx 2k_p$.

## 3. Electrostriction-driven stimulated scattering: hydrodynamic and kinetic regimes

With the pump and Stokes fields phase-matched as needed for the maximum efficiency of stimulated scattering, the coupled equations for $\mathbf{E}_p(\mathbf{r},\mathrm{t})$, $\mathbf{E}_s(\mathbf{r},\mathrm{t})$, and $\delta\rho(\mathbf{r},t)$, solved in the limit of undepleted pump, lead to the signature exponentially growing solution for the Stokes field intensity [2, 3],

$$I_s(z) = \frac{n_s c}{8\pi}\mathbf{E}^*_s\cdot\mathbf{E}_s = I_s(z=0)\exp\left[g_s I_{\mathrm{p}} z\right], \quad (7)$$

with constant pump intensity $I_{\mathrm{p}} = \frac{n_p c}{8\pi}\mathbf{E}^*_p\cdot\mathbf{E}_p$, coordinate $z$ chosen along $\mathbf{k}_s$, gain factor

$$g_s = g_s\left(\mathbf{K},\Omega;\mathbf{e}_p,\mathbf{e}_s\right) = -\frac{32\pi^2\omega_s^2}{n_p k_s c^3}\left|\mathbf{e}^*_s\cdot\mathbf{e}_p\right|^2\mathrm{Im}\left[\chi^{(3)}(\mathbf{K},\Omega)\right], \quad (8)$$

and refractive index at pump frequency $n_p$.

Finding the gain factor $g_s$ requires a specific physical model connecting the field-induced density change $\delta\rho(\mathbf{r},t)$ to thermodynamic parameters of the medium. In standard settings of stimulated Brillouin scattering [2, 3, 5, 11, 12], such a model is formulated in terms of the Navier

– Stokes equation in conjunction with the continuity equation for $\rho(\mathbf{r}, t)$, providing equations of motion for the gas-density and temperature variations. In this canonical picture, SBS is adequately understood as a light–sound interaction in which an intense optical pump couples to a Stokes field via an intermediary of a yoked acoustic wave, leading to Stokes-field amplification.

This picture, however, is not universal. In moderate- or low-pressure gases, SBS-type Stokes amplification is also observed in settings in which the spacing $l_g = 2\pi/K$ of the density grating needed to satisfy momentum conservation $\mathbf{k}_p = \mathbf{k}_s + \mathbf{K}$ is smaller, $l_g < l_0$, than the mean free path $l_0$ of gas particles. Although generation of sound waves with such $K$ is not possible, the Stokes field, as kinetic analysis shows [29, 30], can still be amplified. Such amplification can be understood in terms of stimulated scattering off the light-induced grating of the density of ballistic, free-streaming gas particles. Because $l_0 = v_0/\nu_c$, where is the mean velocity of gas particles and $\nu_c$ is the collision frequency, the ratio $l_g/l_0$ can be expressed as $l_g/l_0 = 2\pi y$, where $y = \nu_c/(K v_0)$. As a useful reference, for a backward stimulated scattering ($\theta = \pi$, $K \approx 2k_p$) of $\lambda_p \approx 266$ nm laser radiation in 1 atm of argon at room temperature, the $y$ parameter is estimated as $y \approx 0.26$, indicating the kinetic regime of light scattering.

## 4. Kinetic framework

In the established framework for the analysis of stimulated and coherent light scattering in this regime [32, 34, 80, 98], the field-induced density variation $\delta\rho_\Omega$ is found by solving a set of kinetic equations for the velocity moments derived as a closure for the Boltzmann equation

$$\left(\frac{\partial}{\partial t} + \mathbf{v} \cdot \boldsymbol{\nabla}\right) f + \frac{\mathbf{F}}{m} \cdot \boldsymbol{\nabla}_{\mathbf{v}} f = \mathcal{C}[f] \qquad (9)$$

for the one-particle distribution function $f \equiv f(\mathbf{r}, \mathbf{v}, t)$ in an ensemble of particles with mass $m$ close to its equilibrium in the presence of the laser-driven ponderomotive force $\mathbf{F} = \mathbf{F}(\mathbf{r}, \mathrm{t})$ [Eq. (3)] with a suitable linearization of the collision operator $\mathcal{C}[f]$.

The solution to Eq. (9) is sought for perturbatively, with $f(\mathbf{r}, \mathbf{v}, t)$ expanded as

$$f(\mathbf{r}, \mathbf{v}, t) = \rho_0 f_0(\mathbf{r}, \mathbf{v}, t) + f_1(\mathbf{r}, \mathbf{v}, t) \qquad (10)$$

about the equilibrium stationary distribution $f_0(\mathbf{r}, \mathbf{v}, t)$ with nonperturbed density $\rho_0$.

Substituting Eq. (10) into Eq. (9) yields a linearized equation for the small correction $f_1 \equiv f_1(\mathbf{r}, \mathbf{v}, t)$,

$$\left(\frac{\partial}{\partial t} + \mathbf{v} \cdot \boldsymbol{\nabla}\right) f_1 + \frac{\rho_0}{m} \mathbf{F} \cdot \boldsymbol{\nabla}_{\mathbf{v}} f_0 = \mathcal{C}_1[f_1], \qquad (11)$$

with $\mathcal{C}_1[f_1] = \mathcal{C}[\rho_0 f_0 + f_1] - \mathcal{C}[\rho_0 f_0]$.

For a Maxwellian equilibrium distribution,

$$f_0(\mathbf{r}, \mathbf{v}, t) = f_0(\mathbf{v}) = \frac{1}{\pi^{3/2} v_0^3} \exp\left(-\frac{v^2}{v_0^2}\right), \tag{12}$$

where $v_0 = \sqrt{2k_B T_0/m}$ is thermal velocity, $T_0$ is the equilibrium temperature, and $k_B$ is the Boltzmann constant, the collision term in Eq. (11), written in the linear BGK approximation, is expressed as

$$\mathcal{C}_1[f_1] = \mathcal{C}_{\mathrm{BGK}}[f_1] = -\tau^{-1}(f_1 - \delta f_M) \tag{13}$$

with

$$\delta f_M = f_0 \left[\delta\rho + \frac{2\rho_0}{v_0^2} \mathbf{u} \cdot \mathbf{v} + \rho_0 \left(\frac{v^2}{v_0^2} - \frac{3}{2}\right) \frac{\delta T}{T_0}\right], \tag{14}$$

relaxation-time parameter $\tau = \nu_c^{-1}$, and first-order corrections to the density, velocity and temperature

$$\delta\rho = \delta\rho(\mathbf{r}, t) = \int f_1(\mathbf{r}, \mathbf{v}, t) d^3 v \tag{15}$$

$$\mathbf{u} = \mathbf{u}(\mathbf{r}, t) = \rho_0^{-1} \int \mathbf{v} f_1(\mathbf{r}, \mathbf{v}, t) d^3 v, \tag{16}$$

$$\delta T = \delta T(\mathbf{r}, t) = \frac{2}{3\rho_0 v_0^2} T_0 \int v^2 f_1(\mathbf{r}, \mathbf{v}, t) d^3 v - \frac{\delta\rho(\mathbf{r}, t)}{\rho_0} T_0. \tag{17}$$

Because

$$\int d^3 v \begin{pmatrix} 1 \\ \mathbf{v} \\ v^2 \end{pmatrix} C_{BGK}^{(1)}[f_1] = \begin{pmatrix} 0 \\ \mathbf{0} \\ 0 \end{pmatrix}, \tag{18}$$

Eq. (11) with a linear BGK collision term $\mathcal{C}_{\mathrm{BGK}}[f_1]$ [Eq. (13)] conserves the number of particles, momentum, and energy. This equation derives directly from the linearized Boltzmann equation [85, 87 – 89, 99], approximating all the nonzero eigenvalues of the collision operator with a single constant $\tau = \nu_c^{-1}$ and is therefore often referred to as the single-relaxation-time model.

With a laser field as defined by Eq. (1), the spectral component of the force $\mathbf{F}(\mathbf{r}, \mathrm{t})$ driving a density grating with frequency $\Omega$ and wave vector $\mathbf{K}$ is found from Eq. (3) as

$$\mathbf{F}_{\mathbf{K},\Omega} = \frac{i\alpha\mathbf{K}}{4} \left(\mathbf{E}_p \cdot \mathbf{E}_s^*\right). \tag{19}$$

For a laser driver (1) with copolarized pump and Stokes fields, the $\mathbf{K}$ projection of this force is

$$\mathbf{F}(\mathbf{r}, \mathrm{t})|_{\mathbf{K},\Omega} = F_\Omega \exp[i(\mathbf{K} \cdot \mathbf{r} - \Omega t + \pi/2)] + \mathrm{c.c.} \tag{20}$$

with $F_\Omega = \alpha K E_p E_s/4$.

Searching for small perturbations $f_1(\mathbf{r},\mathbf{v},t)$ , $\mathbf{u}(\mathbf{r},t)$, and $\delta T(\mathbf{r},t)$ in the form of

$$f_1(\mathbf{r},\mathbf{v},t) = f_{1,\mathbf{K},\Omega}(\mathbf{v})\exp(i\Phi) + f^*_{1,\mathbf{K},\Omega}(\mathbf{v})\exp(-i\Phi), \quad (21)$$

$$\mathbf{u}(\mathbf{r},t) = \mathbf{u}_{\mathbf{K},\Omega}\exp[i(\mathbf{K}\cdot\mathbf{r} - \Omega t)] + \mathbf{u}^*_{\mathbf{K},\Omega}\exp(-i\Phi), \quad (22)$$

$$\delta T(\mathbf{r},t) = \delta T_{\mathbf{K},\Omega}\exp[i(\mathbf{K}\cdot\mathbf{r} - \Omega t)] + \delta T^*_{\mathbf{K},\Omega}\exp(-i\Phi), \quad (23)$$

leads to a set of linear equations for $f_{1,\mathbf{K},\Omega}(\mathbf{v})$, $\delta\rho_{\mathbf{K},\Omega}$, $\mathbf{u}_{\mathbf{K},\Omega}$, and $\delta T_{\mathbf{K},\Omega}$, such that

$$\delta\rho_{\mathbf{K},\Omega} = \int f_{1,\mathbf{K},\Omega}(\mathbf{v})d^3v\,, \quad (24)$$

$$\mathbf{u}_{\mathbf{K},\Omega} = \rho_0^{-1}\int \mathbf{v}\, f_{1,\mathbf{K},\Omega}(\mathbf{v})d^3v, \quad (25)$$

$$\delta T_{\mathbf{K},\Omega} = \frac{2}{3\rho_0 v_0^2}T_0\int v^2 f_{1,\mathbf{K},\Omega}(\mathbf{v})d^3v - \frac{\delta\rho_{\mathbf{K},\Omega}}{\rho_0}T_0. \quad (26)$$

The linearity of these equations is guaranteed by the perturbative procedure that treats $f_1(\mathbf{r},\mathbf{v},t)$ , $\delta\rho(\mathbf{r},t)$, $\mathbf{u}(\mathbf{r},t)$, and $\delta T(\mathbf{r},t)$ as small corrections to the respective equilibrium functions in the first order in the external force **F**. Solving this set of equations for the laser-induced density variation $\delta\rho_{\mathbf{K},\Omega}$ and plugging this solution into Eq. (6) for the Fourier-space nonlinear polarization $\mathbf{P}^{(3)}_{\mathbf{K},\Omega}$, we set up the nonlinear coupling term in equations for the pump, Stokes, and polarization amplitudes [1 – 3, 5 – 7] as a framework to find the gain $g_s(\mathbf{K},\Omega;\mathbf{e}_p,\mathbf{e}_s)$ in stimulated scattering [84, 98, 100] and describe the coherent signal in coherent scattering from density gratings [32 – 34, 80].

## 5. Fluctuation – dissipation inner workings behind electrostriction-driven stimulated light scattering

Here, in search of deeper insights into the inner workings of nonlinear light scattering in kinetic settings, we explore the properties of such processes as imposed by the fluctuation – dissipation theorem. Working toward this goal, we identify the stimulated gain/loss in such a process as the amplification/dissipation counterpart of spontaneous density fluctuations. This relation, as we show below, connects the stimulated gain/loss spectrum to the spectrum of spontaneous **K**-mode density fluctuations $S_{\mathbf{K}}(\Omega)$, allowing the stimulated scattering gain $g_s(\mathbf{K},\Omega;\mathbf{e}_p,\mathbf{e}_s)$ to be found directly from $S_{\mathbf{K}}(\Omega)$, thus making the solution of kinetic equations with the field-driven electrostriction term unnecessary.

The realization that the response of a dynamic system to an external force is intrinsically related to spontaneous fluctuations in this system is a fundamental result of statistical thermodynamics, broadly referred to as the fluctuation – dissipation theorem (FDT). Early FDT precursors are found in the Einstein theory of Brownian motion [101] and Nyquist's analysis of noise in electric currents [102]. The formal proof of the FDT has been provided by Callen and Welton [103] and expanded by Kubo [104]. While the power of the FDT as an analytical tool applicable to a vast variety of physical settings is broadly appreciated [105 – 110], accurately isolating the fluctuation and dissipation properties that connect via the FDT, thus enabling the FDT-based machinery, is often nontrivial, especially for nonlinear processes, such as stimulated and coherent light scattering.

Aiming to provide grounds for consistent and rigorous FDT-based treatment, we seek to identify the Hamiltonian properties of laser-driven electrostriction as a source of stimulated and coherent light scattering. Working to this end, we observe that the electrostriction-induced correction to dielectric permittivity (4) translates into a change in energy density [3, 6, 7]

$$\delta u(\mathbf{r},\mathrm{t}) = -\frac{1}{8\pi}\left(\frac{\partial\varepsilon}{\partial\rho}\right)[\mathbf{E}(\mathbf{r},\mathrm{t})]^2\delta\rho(\mathbf{r},\mathrm{t}) = -\hbar(\mathbf{r},\mathrm{t})\delta\rho(\mathbf{r},\mathrm{t}), \tag{27}$$

With

$$\hbar(\mathbf{r},\mathrm{t}) = -(8\pi)^{-1}(\partial\varepsilon/\partial\rho)[\mathbf{E}(\mathbf{r},\mathrm{t})]^2, \tag{28}$$

connecting to the ponderomotive force $\mathbf{F}(\mathbf{r},\mathrm{t})$ [Eq. (3)] via $\mathbf{F}(\mathbf{r},\mathrm{t}) = m\nabla\hbar(\mathbf{r},\mathrm{t})$.

Treating $\delta u(\mathbf{r},\mathrm{t})$ in Eq. (27) as the density of the interaction Hamiltonian and observing that $\hbar(\mathbf{r},\mathrm{t}) = -\partial(\delta u)/\partial(\delta\rho)$, we see that, in the Hamiltonian picture of such interaction, $\hbar(\mathbf{r},\mathrm{t})$ is coupled to $\delta\rho(\mathbf{r},\mathrm{t})$ as its canonical conjugate. The interaction Hamiltonian in this picture is found via space integration of the Hamiltonian density $\delta u(\mathbf{r},\mathrm{t})$,

$$H_{\mathrm{int}}(t) = \int_V \delta u(\mathbf{r},\mathrm{t})\, d^3r = -\int_V \hbar(\mathbf{r},t)\,\delta\rho(\mathbf{r},\mathrm{t})d^3r. \tag{29}$$

With $\delta\rho(\mathbf{r},\mathrm{t})$ and $\hbar(\mathbf{r},\mathrm{t})$ expanded in plane-wave modes as

$$\delta\rho(\mathbf{r},t) = \sum_{\mathbf{K}}\delta\rho_{\mathbf{K}}(t)\exp(i\mathbf{K}\cdot\mathbf{r}) \tag{30}$$

and

$$\hbar(\mathbf{r},t) = \sum_{\mathbf{K}}\hbar_{\mathbf{K}}(t)\exp(i\mathbf{K}\cdot\mathbf{r}), \tag{31}$$

where

$$\delta\rho_{\mathbf{K}}(t) = V^{-1}\int_V \delta\rho(\mathbf{r},t)\exp(-i\mathbf{K}\cdot\mathbf{r})\, d^3r \tag{32}$$

and

$$\hbar_{\mathbf{K}}(t) = V^{-1}\int_V \hbar(\mathbf{r},t)\exp(-i\mathbf{K}\cdot\mathbf{r})\, d^3r, \tag{33}$$

such that $\delta\rho_{-\mathbf{K}} = \delta\rho_{\mathbf{K}}^*$ and $\hbar_{-\mathbf{K}} = \hbar_{\mathbf{K}}^*$, the interaction Hamiltonian in Eq. (29) is

$$H_{\text{int}}(t) = -V \sum_{\mathbf{K}} \hbar_{-\mathbf{K}}(t)\delta\rho_{\mathbf{K}}(t). \tag{34}$$

The effect of electrostriction can thus be understood in terms of the Liouville equation for the phase-space density $q(t)$,

$$\frac{\partial q(t)}{\partial t} = \mathcal{L}q(t), \tag{35}$$

with the Liouville operator $\mathcal{L}$ defined as

$$\mathcal{L}q(t) = -\{q(t), H(t)\}, \tag{36}$$

where $H(t)$ is the full evolution Hamiltonian and $\{\cdot,\cdot\}$ is the Poisson bracket over the phase-space coordinates.

With the evolution Hamiltonian $H(t)$ given by the sum

$$H(t) = H_0 + H_{\text{int}}(t) \tag{37}$$

of the unperturbed, field-free Hamiltonian $H_0$ and the interaction Hamiltonian $H_{\text{int}}(t)$, Eq. (36) gives

$$\mathcal{L}q(t) = (\mathcal{L}_0 + \mathcal{L}_{\text{int}})q(t) = \{H_0, q(t)\} + \{H_{\text{int}}(t), q(t)\}. \tag{38}$$

Solving Eq. (35) to the first order in $\hbar$ subject to the initial condition

$$q(t \to -\infty) = q_{\text{eq}} = Z^{-1}\exp(-\beta H_0), \tag{39}$$

where $\beta = (k_B T)^{-1}$, $T$ is the temperature, $k_B$ is the Boltzmann constant, and $Z$ is the canonical partition function, and performing integration by parts in the phase space, we find

$$q(t) = q_{\text{eq}} + \Delta q(t), \tag{40}$$

with

$$\Delta q(t) = \int_{-\infty}^{t} \exp[\mathcal{L}_0(t - t')]\,\mathcal{L}_{\text{int}}(t')q_{\text{eq}}dt'. \tag{41}$$

The change in the expectation value of $\delta\rho_{\mathbf{K}}(t)$ induced by the light – matter interaction governed by the Hamiltonian (34) is found as

$$\langle\delta\rho_{\mathbf{K}}(t)\rangle_{\text{int}} = \int \Delta q(t)\delta\rho_{\mathbf{K}}d\xi\,, \tag{42}$$

where integration is over the entire phase space of $q(t)$.

Combining Eq. (42) with Eqs. (30) – (34) yields

$$\langle\delta\rho_{\mathbf{K}}(t)\rangle_{\text{int}} = V\int_{-\infty}^{t} dt'\,\hbar_{\mathbf{K}}(t')\int d\xi\delta\rho_{\mathbf{K}}(t - t')\{q_{\text{eq}}, \delta\rho_{-\mathbf{K}}(0)\}. \tag{43}$$

With the phase-space integral in Eq. (43) taken by parts, the result can be expressed as the expectation value with respect to the equilibrium density $q_{\text{eq}}$,

$$\int \delta\rho_{\mathbf{K}}(t)\{q_{\text{eq}}, \delta\rho_{-\mathbf{K}}(0)\}d\xi = \langle\{\delta\rho_{-\mathbf{K}}(0), \delta\rho_{\mathbf{K}}(t)\}\rangle_{\text{eq}}, \tag{44}$$

leading to

$$\langle\delta\rho_{\mathbf{K}}(t)\rangle_{\text{int}} = V\int_{-\infty}^{t} \hbar_{\mathbf{K}}(t')\langle\{\delta\rho_{-\mathbf{K}}(0), \delta\rho_{\mathbf{K}}(t-t')\}\rangle_{\text{eq}}dt' \,. \tag{45}$$

Eq. (44) is recognized as an expression of $\langle\delta\rho_{\mathbf{K}}(t)\rangle_{\text{int}}$ in terms of causal linear response to an external driver $\hbar_{\mathbf{K}}(t)$,

$$\langle\delta\rho_{\mathbf{K}}(t)\rangle_{\text{int}} = \int_{-\infty}^{t} Q_{\mathbf{K}}(t-t')\hbar_{\mathbf{K}}(t')dt', \tag{46}$$

with

$$Q_{\mathbf{K}}(t) = V\theta(t)\langle\{\delta\rho_{-\mathbf{K}}(0), \delta\rho_{\mathbf{K}}(t)\}\rangle_{\text{eq}}, \tag{47}$$

where $\theta(t)$ is the unit step function.

Fourier transform of Eq. (46) gives

$$\langle\delta\rho_{\mathbf{K}}(\Omega)\rangle_{\text{int}} = Q_{\mathbf{K}}(\Omega)\hbar_{\mathbf{K}}(\Omega), \tag{48}$$

with $\delta\rho_{\mathbf{K}}(\Omega) = \int_{-\infty}^{\infty} \delta\rho_{\mathbf{K}}(t)e^{i\Omega t}dt$, $Q_{\mathbf{K}}(\Omega) = \int_{-\infty}^{\infty} Q_{\mathbf{K}}(t)e^{i\Omega t}dt$, and $\hbar_{\mathbf{K}}(\Omega) = \int_{-\infty}^{\infty} \hbar_{\mathbf{K}}(t)e^{i\Omega t}dt$.

We now recognize $\langle\delta\rho_{\mathbf{K}}(t)\delta\rho_{-\mathbf{K}}(0)\rangle_{\text{eq}}$ as the equilibrium autocorrelation function [104 – 107],

$$\mathcal{R}_{\mathbf{K}}(t) = \langle\delta\rho_{\mathbf{K}}(t)\delta\rho_{-\mathbf{K}}(0)\rangle_{\text{eq}}. \tag{49}$$

Because $\delta\rho(t)$ at equilibrium is stationary, $\mathcal{R}_{\mathbf{K}}(-t) = \mathcal{R}_{\mathbf{K}}^{*}(t)$, differentiation of $\mathcal{R}_{\mathbf{K}}(t)$ in $t$ gives

$$\frac{d\mathcal{R}_{\mathbf{K}}(t)}{dt} = -\left\langle\dot{\delta\rho}_{-\mathbf{K}}(0)\delta\rho_{\mathbf{K}}(t)\right\rangle_{\text{eq}}. \tag{50}$$

Combining Eqs. (39), (47) – (50), we find

$$Q_{\mathbf{K}}(t) = -\beta V\theta(t)\frac{d\mathcal{R}_{\mathbf{K}}(t)}{dt}. \tag{51}$$

Fourier transform of Eq. (51) yields

$$Q_{\mathbf{K}}(\Omega) = -\beta V\int_{0}^{\infty}\frac{d\mathcal{R}_{\mathbf{K}}(t)}{dt}\exp(i\Omega t)\,dt. \tag{52}$$

Taking the integral on the right-hand side of Eq. (52) by parts with a standard requirement that spontaneous fluctuations should decay in the $t \to \infty$ limit, we find

$$Q_{\mathbf{K}}''(\Omega) \equiv \text{Im}Q_{\mathbf{K}}(\Omega) = \pi\beta m\rho_0\Omega S_{\mathbf{K}}(\Omega), \tag{53}$$

where

$$S_{\mathbf{K}}(\Omega) = \frac{S_{\mathbf{K}}^{\text{raw}}(\Omega)}{2\pi\mathcal{R}_{\mathbf{K}}(0)} \tag{54}$$

and

$$S_{\mathbf{K}}^{\mathrm{raw}}(\Omega) = \int_{-\infty}^{\infty} \mathcal{R}_{\mathbf{K}}(t) \exp(i\Omega t)\, dt \tag{55}$$

is the raw two-sided density spectrum defined as the Fourier transform of the equilibrium autocorrelation $\mathcal{R}_{\mathbf{K}}(t)$. $S_{\mathbf{K}}(\Omega)$ as defined by Eqs. (54) and (55) is thus the normalized spectrum of equilibrium density fluctuations, $\int_{-\infty}^{\infty} S_{\mathbf{K}}(\Omega) d\Omega = 1$, measured in units of time, which connects to the spectrum $Q_{\mathbf{K}}(\Omega)$ of the linear response to the electrostrictive driver $\hbar_{\mathbf{K}}(t)$ via the FDT-type relation of Eq. (53).

Eqs. (51) and (53) express fluctuation – dissipation relations for a fluid driven by an electrostrictive force. Specifically, Eq. (51) relates the **K** mode of the linear response $Q_{\mathbf{K}}(t)$ of the fluid to the electrostriction-induced force to the equilibrium autocorrelation function $\mathcal{R}_{\mathbf{K}}(t) = \langle \delta\rho_{\mathbf{K}}(t)\delta\rho_{-\mathbf{K}}(0)\rangle_{\mathrm{eq}}$ of density fluctuations. Eq. (53), on the other hand, expresses the fluctuation – dissipation connection between the imaginary part of the spectrum $Q_{\mathbf{K}}(\Omega)$ of the response $Q_{\mathbf{K}}(t)$ and the spectrum $S_{\mathbf{K}}(\Omega)$ of spontaneous fluctuations in this fluid. In physical terms, these relations establish that the dynamics of disturbances induced in a medium by an external electrostriction-induced force unfolds via the same pathways as the dynamics of internal, spontaneous fluctuations in this medium at equilibrium, e.g., fluctuations launched by the initial condition $f_0(\mathbf{r}, \mathbf{v}, t = 0) = f_0(\mathbf{v})$ for the one-particle distribution function $f(\mathbf{r}, \mathbf{v}, t)$ obeying Eqs. (9) – (14). Eq. (51) can be viewed as an expression of the fluctuation – dissipation theorem in the form of Onsager's regression principle [111, 112], as it relates the time-derivative of the autocorrelation of spontaneous density fluctuations to the linear response to an external force, thus establishing that the relaxation of a macroscopic system perturbed out of equilibrium follows the same laws as the decay of spontaneous fluctuations at equilibrium.

To understand connection between fluctuations and dissipation in stimulated and coherent light scattering, we observe from Eqs. (19), (28), and (48) that the amplitude of the $(\mathbf{K}, \Omega)$ mode of density modulation driven by the laser field Eq. (1) is

$$\delta\rho_{\mathbf{K},\Omega} = \frac{\gamma_e}{16\pi\rho_0} Q_{\mathbf{K}}(\Omega)\left(\mathbf{E}_p \cdot \mathbf{E}_s^*\right) = \frac{\alpha}{4m} Q_{\mathbf{K}}(\Omega)\left(\mathbf{E}_p \cdot \mathbf{E}_s^*\right). \tag{56}$$

Substituting Eq. (56) into Eq. (6) gives

$$\mathbf{P}_{\mathbf{K},\Omega}^{(3)} = \frac{\gamma_e^2}{64\pi^2\rho_0^2} Q_{\mathbf{K}}^*(\Omega)\mathbf{E}_p\left(\mathbf{E}_p^* \cdot \mathbf{E}_s\right). \tag{57}$$

The third-order nonlinear susceptibility $\chi^{(3)}(\mathbf{K}, \Omega)$ is thus expressed as

$$\chi^{(3)}(\mathbf{K}, \Omega) = \frac{\gamma_e^2}{64\pi^2\rho_0^2} Q_{\mathbf{K}}^*(\Omega). \tag{58}$$

The stimulated gain [Eq. (8)] is now found as

$$g_s(\mathbf{K},\Omega;\mathbf{e}_p,\mathbf{e}_s)=\frac{\gamma_e^2\omega_s^2}{2n_pk_sc^3\rho_0^2}\left|\mathbf{e}_s^*\cdot\mathbf{e}_p\right|^2\mathrm{Im}[Q_{\mathbf{K}}(\Omega)]. \tag{59}$$

Applying the FDT in the form of Eq. (53) gives

$$g_s(\mathbf{K},\Omega;\mathbf{e}_p,\mathbf{e}_s)=\frac{\pi\gamma_e^2\omega_s}{n_pn_sc^2\rho_0v_0^2}\left|\mathbf{e}_s^*\cdot\mathbf{e}_p\right|^2\Omega S_{\mathbf{K}}(\Omega). \tag{60}$$

As a meaningful units-of-measure check, $\hbar_{\mathbf{K}}$ in Eqs. (31), (34), (45), (46) is in units of (length/time)$^2$, or energy/mass, such that the Hamiltonian $H_{\mathrm{int}}$, defined as the volume integral of the energy density [Eq. (29)], has the units of energy, and the ponderomotive force, $\mathbf{F}=m\nabla\hbar$, is in units of mass×length/time$^2$, i.e., mass×acceleration. The density mode correlation $\mathcal{R}_{\mathbf{K}}(t)$ in Eqs. (49) – (52) is in units of mass$^2$/length$^6$. The response $Q_{\mathbf{K}}$ in Eqs. (48), (52), (53), and (56) – (59) is in units of mass$^2$/(energy×length$^3$), while $S_{\mathbf{K}}(\Omega)$ in Eq. (53) is in units of inverse time, satisfying the sum rule, $\int_{-\infty}^{\infty}S_{\mathbf{K}}(\Omega)d\Omega=1$. The stimulated gain in Eqs. (59) and (60) is in units of length × time/energy, such that the product $(g_sI_{\mathrm{p}})^{-1}$, which defines the gain buildup length, with field intensity [Eq. (7)] in units of energy/(time×length$^2$) is in units of length. With $\omega_s$, $c$, and $\rho_0v_0^2$ on the right-hand side of Eq. (60) carrying the units of 1/time, length/time, and energy/length$^3$, the factor $\frac{\pi\gamma_e^2\omega_s}{n_pn_sc^2\rho_0v_0^2}$ is in units of length × time/energy, as it is required for correctly defined $g_s$.

Eq. (60) connects the gain that the Stokes field experiences as a part of such nonlinear process to spontaneous fluctuations inside the medium. As one important insight, Eq. (60) shows that the solution of the kinetic equation with an external field [Eq. (9) or (11)] is not needed for the analysis of stimulated gain and loss as all the physics behind stimulated gain and loss is fully captured by a simpler equation – the respective basic kinetic equation without an external field. All the parameters of stimulated scattering, including the stimulated gain and loss, the spectrum of the gain spectrum, location of its maximum, and its bandwidth, can also be found directly from the spectrum of spontaneous density fluctuations, which, in its turn, is found by solving the suitable kinetic equation for the system without an external field, e.g., the Boltzmann equation with the linearized BGK collision operator without an external force term [Eqs. (9), (11) with $\mathbf{F}=0$].

In the following sections, we will demonstrate that the FDT solution for the stimulated gain found in Eq. (60) allows an accurate description of stimulated light scattering within a vast parameter space, recovering all the signature properties of sound-wave-mediated Stokes amplification in the hydrodynamic regime, providing a continuous crossover from the hydrodynamic to kinetic regime of stimulated scattering, and explaining distinctly different properties of kinetic stimulated scattering from density gratings, consistent with experiments on stimulated scattering in moderate-pressure gases.

## 6. Understanding the stimulated gain

### *6.1. General solution*

To set up a benchmark for the assessment of the adequacy and accuracy of the kinetic and hydrodynamic approximations for the description of stimulated light scattering, we operate within the framework of the Boltzmann equation with a linearized BGK collision term $\mathcal{C}_{\mathrm{BGK}}[f_1]$ [Eqs. (13), (14)], searching for the solution for the distribution function $f(\mathbf{r},\mathbf{v},t)$ in the form of Eq. (10). As one of the central steps toward this goal, we substitute Eqs. (13, (14), (21) – (23) into Eq. (11) to arrive at

$$(\nu_c - i\Omega + iKv_{\parallel})f_{1,\mathbf{K},\Omega} = \nu_c \delta f_M + \frac{2iK\rho_0 \hbar_{\mathbf{K},\Omega} v_{\parallel}}{v_0^2} f_0, \tag{61}$$

where $v_{\parallel} = \mathbf{v}\cdot\mathbf{K}/K$.

Solving this equation for $f_{1,\mathbf{K},\Omega}$, we find

$$f_{1,\mathbf{K},\Omega}(\mathbf{v}) = \frac{2\mathcal{w}\mathcal{v} - iy[\mathcal{r} + 2\mathcal{u}\mathcal{v} + \left(\mathcal{v}^2 + s - \frac{3}{2}\right)\Theta]}{\mathcal{v} - \zeta}\rho_0 f_0(\mathbf{v}), \tag{62}$$

with $y = \frac{\nu_c}{Kv_0}$, $\mathcal{v} = \frac{v_{\parallel}}{v_0}$, $\mathcal{r} = \frac{\delta\rho_{\mathbf{K},\Omega}}{\rho_0}$, $\mathcal{u} = \frac{u_{\parallel,\mathbf{K},\Omega}}{v_0}$, $\mathcal{w} = \frac{\hbar_{\mathbf{K},\Omega}}{v_0^2}$, $s = \frac{v_\perp^2}{v_0^2}$, $\Theta = \frac{\delta T_{\mathbf{K},\Omega}}{T_0}$, $\mathbf{v}_\perp = \mathbf{v} - v_{\parallel}\,\mathbf{K}/K$, $\zeta = x + iy$, $x = \frac{\Omega}{Kv_0}$, and $u_{\parallel,\mathbf{K},\Omega} = \rho_0^{-1}\int v_{\parallel}\, f_{1,\mathbf{K},\Omega}(\mathbf{v})d^3v$.

We now observe that the zeroth-moment integral $\int\{\cdot\}d^3v$ of $f_{1,\mathbf{K},\Omega}(\mathbf{v})$ on the left-hand side of Eq. (62) gives $\int f_{1,\mathbf{K},\Omega}(\mathbf{v})d^3v = \delta\rho_{\mathbf{K},\Omega} = \rho_0\mathcal{r}$. Taking the $\int\{\cdot\}d^3v$ integral of the entire Eq. (62), we find

$$\mathcal{r} + iy[J_0\mathcal{r} + 2J_1\mathcal{u} + \left(J_2 - \tfrac{1}{2}J_0\right)\Theta] = 2\mathcal{w}J_1, \tag{63}$$

with

$$J_n(\zeta) = \frac{1}{\sqrt{\pi}}\int_{-\infty}^{\infty}\frac{t^n e^{-t^2}}{t-\zeta}\,dt, \tag{64}$$

such that $J_0 = i\sqrt{\pi}\,w(\zeta)$, where $w(\zeta) = e^{-\zeta^2}\mathrm{erfc}(-i\zeta)$ is the Faddeeva function [113].

The first-moment integral $\int v_{\parallel}\{\cdot\}d^3v$ of $f_{1,\mathbf{K},\Omega}(\mathbf{v})$, in its turn, yields $\int v_{\parallel}\, f_{1,\mathbf{K},\Omega}(\mathbf{v})d^3v = \rho_0 u_{\parallel,\mathbf{K},\Omega}$. $\int v_{\parallel}\{\cdot\}d^3v$ integration of Eq. (62) thus leads to

$$u + iy[J_1 r + 2J_2 u + \left(J_3 - \tfrac{1}{2}J_1\right)\Theta] = 2wJ_2. \tag{65}$$

The second-moment integral $\int v^2\{\cdot\}d^3v$ of $f_{1,\mathbf{K},\Omega}(\mathbf{v})$ is $\int v^2 f_{1,\mathbf{K},\Omega}(\mathbf{v})d^3v = \frac{3}{2}\rho_0 v_0^2\left(\frac{\delta T_{\mathbf{K},\Omega}}{T_0} + \frac{\delta\rho_{\mathbf{K},\Omega}}{\rho_0}\right)$. Taking $\int v^2\{\cdot\}d^3v$ of Eq. (62), we arrive at

$$\Theta + iy[\tfrac{2}{3}\left(J_2 - \tfrac{1}{2}J_0\right) r + \tfrac{4}{3}\left(J_3 - \tfrac{1}{2}J_1\right) u + \tfrac{2}{3}\left(J_4 - J_2 + \tfrac{5}{4}J_0\right)\Theta] = \tfrac{4}{3}w\left(J_3 - \tfrac{1}{2}J_1\right). \tag{66}$$

The set of equations (63), (65), and (66) can be written in the matrix form as

$$[\mathbf{I} + iy\mathbf{M}_J(\zeta)]\mathbf{m} = w\,\mathbf{c}_J(\zeta), \tag{67}$$

where

$$\mathbf{m} = \begin{pmatrix} r \\ u \\ \Theta \end{pmatrix} = \begin{pmatrix} \delta\rho_{\mathbf{K},\Omega}/\rho_0 \\ u_{\parallel,\mathbf{K},\Omega}/v_0 \\ \delta T_{\mathbf{K},\Omega}/T_0 \end{pmatrix}, \tag{68}$$

$$\mathbf{M}_J = \begin{pmatrix} J_0 & 2J_1 & J_2 - \frac{1}{2}J_0 \\ J_1 & 2J_2 & J_3 - \frac{1}{2}J_1 \\ \frac{2}{3}\left(J_2 - \frac{1}{2}J_0\right) & \frac{4}{3}\left(J_3 - \frac{1}{2}J_1\right) & \frac{2}{3}\left(J_4 - J_2 + \frac{5}{4}J_0\right) \end{pmatrix}, \tag{69}$$

$$\mathbf{c}_J = 2\begin{pmatrix} J_1 \\ J_2 \\ \frac{2}{3}\left(J_3 - \frac{1}{2}J_1\right) \end{pmatrix}, \tag{70}$$

and $\mathbf{I}$ is the $3 \times 3$ identity matrix.

Solving Eq. (67) for $\mathbf{m}$ gives

$$\mathbf{m} = w\,[\mathbf{I} + iy\mathbf{M}_J(\zeta)]^{-1}\mathbf{c}_J = w\,\boldsymbol{\mathcal{A}}^{-1}\mathbf{c}_J, \tag{71}$$

with

$$\boldsymbol{\mathcal{A}} = \mathbf{I} + iy\mathbf{M}_J = \begin{pmatrix} 1 + iyJ_0 & 2iyJ_1 & iy\left(J_2 - \frac{1}{2}J_0\right) \\ iyJ_1 & 1 + 2iyJ_2 & iy\left(J_3 - \frac{1}{2}J_1\right) \\ \frac{2i}{3}y\left(J_2 - \frac{1}{2}J_0\right) & \frac{4i}{3}y\left(J_3 - \frac{1}{2}J_1\right) & 1 + \frac{2i}{3}y\left(J_4 - J_2 + \frac{5}{4}J_0\right) \end{pmatrix}. \tag{72}$$

The density perturbation is found as the first element of $\mathbf{m}$,

$$r = \frac{\delta\rho_{\mathbf{K},\Omega}}{\rho_0} = \frac{h_{\mathbf{K},\Omega}}{v_0^2}\,\mathbf{e}_1[\mathbf{I} + iy\mathbf{M}_J(\zeta)]^{-1}\mathbf{c}_J = \frac{\rho_0}{v_0^2}\frac{\det\boldsymbol{\mathcal{A}}_1(x,y)}{\det\boldsymbol{\mathcal{A}}(x,y)}, \tag{73}$$

where $\mathbf{e}_1 = (1 \quad 0 \quad 0)$ and $\boldsymbol{\mathcal{A}}_1$ is the matrix obtained from $\boldsymbol{\mathcal{A}}$ by replacing the first column of $\boldsymbol{\mathcal{A}}$ by $\mathbf{c}_J$,

$$\boldsymbol{\mathcal{A}}_1 = \begin{pmatrix} 2J_1 & 2iyJ_1 & iy\left(J_2 - \frac{1}{2}J_0\right) \\ 2J_2 & 1 + 2iyJ_2 & iy\left(J_3 - \frac{1}{2}J_1\right) \\ \frac{4}{3}\left(J_3 - \frac{1}{2}J_1\right) & \frac{4i}{3}y\left(J_3 - \frac{1}{2}J_1\right) & 1 + \frac{2i}{3}y\left(J_4 - J_2 + \frac{5}{4}J_0\right) \end{pmatrix}. \quad (74)$$

Combining Eq. (73) with Eq. (48), we find

$$Q_{\mathbf{K}}(\Omega) = \frac{\rho_0}{v_0^2}\,\mathbf{e}_1\left(\mathbf{I} + iy\mathbf{M}_J\right)^{-1}\mathbf{c}_J = \frac{\rho_0}{v_0^2}\frac{\det \boldsymbol{\mathcal{A}}_1(x,y)}{\det \boldsymbol{\mathcal{A}}\,(x,y)} \quad , \quad (75)$$

Substituting Eq. (75) into Eq. (59) for the stimulated gain gives

$$g_s\left(\mathbf{K}, \Omega; \mathbf{e}_p, \mathbf{e}_s\right) = \frac{\gamma_e^2\omega_s^2}{2n_pk_sc^3\rho_0v_0^2}\left|\mathbf{e}_s^* \cdot \mathbf{e}_p\right|^2 \mathrm{Im}\left[\mathbf{e}_1\boldsymbol{\mathcal{A}}^{-1}\mathbf{c}_J\right]$$

$$= \frac{\gamma_e^2\omega_s^2}{2n_pk_sc^3\rho_0v_0^2}\left|\mathbf{e}_s^* \cdot \mathbf{e}_p\right|^2 \mathrm{Im}\left[\frac{\det \boldsymbol{\mathcal{A}}_1(x,y)}{\det \boldsymbol{\mathcal{A}}\,(x,y)}\right] \quad . \quad (76)$$

Eq. (76) fulfils the task of expressing the stimulated gain as a closed-form function of laser – matter interaction parameters. To better understand the properties of this solution, we isolate the real and imaginary parts of the Faddeeva function,

$$U(x,y) = \sqrt{\pi}\ \mathrm{Re}\, w\,(x + iy), \quad (77)$$

$$V(x,y) = \sqrt{\pi}\ \mathrm{Im}\, w\,(x + iy), \quad (78)$$

and express the ratio of determinants in Eqs. (73), (75), and (76) as

$$\frac{\det \boldsymbol{\mathcal{A}}_1(x,y)}{\det \boldsymbol{\mathcal{A}}(x,y)} = 2\frac{\mathcal{g}_1 + i\mathcal{g}_2}{\mathcal{g}_3 + i\mathcal{g}_4}, \quad (79)$$

with

$$\mathcal{g}_1(x,y) = 6 + 2y^2 - x(6 + 4y^2)V + y(2x^2 - 2y^2 - 11)U + 8xyUV + 4y^2(U^2 - V^2), \quad (80)$$

$$\mathcal{g}_2(x,y) = -2xy + 2x(2y^2 + 3)U + y(2x^2 - 2y^2 - 11)V + 8y^2UV + 4xy(V^2 - U^2), \quad (81)$$

$$\mathcal{g}_3(x,y) = 6 + 2y^2 - 4x^2y^2 + y(-4x^4 + 4x^2y^2 - 8x^2 - 2y^2 - 11)U + y(-4x^4 + 4x^2y^2 - 8x^2 - 2y^2 - 11)U + 4y^2(2x^2 + 1)(U^2 - V^2) - 16xy^3UV, \quad (82)$$

$$\mathcal{g}_4(x,y) = 2xy(2x^2 + 5) - 8xy^2(x^2 + 2)U + y(-4x^4 + 4x^2y^2 - 8x^2 - 2y^2 - 11)V + 8y^2(2x^2 + 1)UV + 8xy^3(U^2 - V^2). \quad (83)$$

The imaginary part of the ratio of determinants in Eqs. (73), (75), and (76) is found via the identity

$$\mathrm{Im}\left(\frac{g_1+ig_2}{g_3+ig_4}\right)=\frac{g_2g_3-g_1g_4}{g_3^2+g_4^2}. \tag{84}$$

The stimulated gain can thus be expressed as a rational function of $x=\frac{\Omega}{Kv_0}$ and $y=\frac{\nu_c}{Kv_0}$,

$$g_s(x,y;\mathbf{e}_p,\mathbf{e}_s)=\frac{\gamma_e^2\omega_s^2}{n_pk_sc^3\rho_0v_0^2}\left|\mathbf{e}_s^*\cdot\mathbf{e}_p\right|^2\frac{g_2g_3-g_1g_4}{g_3^2+g_4^2}, \tag{85}$$

where $\mathbf{e}_p$ and $\mathbf{e}_s$ are the unit polarization vectors of the pump and Stokes fields.

### *6.2. Kinetic and hydrodynamic limits*

In the collisionless limit, $\nu_c\to 0$, we find $y\to 0$, entailing $U(x,y)\to U(x,0)=\sqrt{\pi}\exp(-x^2)$, $g_l(x,y)\to g_l(x,0)$, $l=1,2,3,4$, with $g_1(x,0)=6(1-xV)$, $g_2(x,0)=6xU$, $g_3(x,0)=6$, and $g_4(x,0)=0$. The stimulated gain in this regime is

$$g_s(\mathbf{K},\Omega;\mathbf{e}_p,\mathbf{e}_s)=\frac{\sqrt{\pi}\gamma_e^2\omega_s^2}{n_pk_sc^3\rho_0v_0^2}\left|\mathbf{e}_s^*\cdot\mathbf{e}_p\right|^2\frac{\Omega}{Kv_0}\exp\left[-\left(\frac{\Omega}{Kv_0}\right)^2\right]. \tag{86}$$

In the opposite, hydrodynamic limit, $y\gg 1$, a power-series expansion of $U(x,y)$ and $V(x,y)$ in $\varrho=y^{-1}\ll 1$ yields

$$U(x,y)\approx\varrho-\left(x^2+\frac{1}{2}\right)\varrho^3+\left(x^4+3x^2+\frac{3}{4}\right)\varrho^5, \tag{87}$$

$$V(x,y)\approx x\varrho^2-\left(x^3+\frac{32}{x}\right)\varrho^4+\left(x^5+5x^3+\frac{15}{4}x\right)\varrho^6. \tag{88}$$

The stimulated gain in this regime is found as

$$g_s(x,y;\mathbf{e}_p,\mathbf{e}_s)=\frac{\gamma_e^2\omega_s^2}{2n_pk_sc^3\rho_0v_0^2}\left|\mathbf{e}_s^*\cdot\mathbf{e}_p\right|^2\mathcal{F}(x,y), \tag{89}$$

with a spectral form factor

$$\mathcal{F}(x,y)=\frac{\frac{x}{y}\left(\frac{5}{18}+\frac{2x^2}{3}\right)}{x^2\left(\frac{5}{6}-x^2\right)^2+\frac{1}{4y^2}\left(\frac{5}{6}-3x^2\right)^2}, \tag{90}$$

Recognizing $c_0=\sqrt{5/6}$ as the adiabatic sound speed $c_s$ expressed in units of the thermal speed, $c_s=c_0v_0$, we express the form factor $\mathcal{F}(x,y)$ in Eq. (89) as

$$\mathcal{F}(x,y)=\frac{\frac{x}{y}\left(\frac{5}{18}+\frac{2x^2}{3}\right)}{x^2(c_0^2-x^2)^2+\frac{1}{4y^2}(c_0^2-3x^2)^2}. \tag{91}$$

When the phase speed of the laser-driven density grating $v_g = \Omega/K$ is close to the adiabatic sound speed $c_s$, i.e., $x = \frac{\Omega}{(Kv_0)} = c_0 + \xi$, a small-$\xi$ power-series expansion of the numerator and denominator in $\mathcal{F}(x, y)$ about $x = c_0$ yields

$$\mathcal{F}(\xi, y) \approx \frac{c_0}{y(4\xi^2 + y^{-2})}. \tag{92}$$

Expressing Eq. (92) as

$$\mathcal{F}(x, y) \approx \frac{1}{2c_0} \frac{\delta}{(x - c_0)^2 + \delta^2}, \tag{93}$$

with $\delta = y^{-1}$, we readily recognize $\mathcal{F}(x, y)$ as a Lorentzian lineshape, centered at $x = c_0$ with a full width at half-maximum (FWHM) $\delta = y^{-1}$, thus recovering the canonical picture of SBS in elementary nonlinear optics [2, 3, 5, 11].

## 7. Stimulated gain from fluctuation – dissipation relation

In Figs. 1a and 1b, we show the spectrum $S_K(\Omega)$ of the equilibrium correlation function $\mathcal{R}_{\mathbf{K}}(t) = \langle \delta\rho_{\mathbf{K}}(t)\delta\rho_{-\mathbf{K}}(0)\rangle_{\mathrm{eq}}$ of spontaneous density fluctuations, calculated for different values of the $y$ parameter using the Yip – Nelkin (YN) solution [85, 88] to the Boltzmann equation with the linearized BGK collision term without an external force, i.e., Eqs. (11) – (14) with $\mathbf{F} = 0$. As one of their most prominent features, large-$y$ density fluctuation spectra are seen to display well-resolved Lorenztian-shaped Brillouin sidebands (Fig. 1b), arising from acoustic modes of density fluctuations, displaced symmetrically about the central peak, corresponding to Rayleigh scattering, caused by thermal fluctuations. This Rayleigh – Brillouin triplet is well-known in spectroscopy of light scattering and is broadly used for the diagnostics of gas flows and analysis of fluid dynamics [31 – 34, 78 – 83].

As $y$ decreases, the Brillouin sidebands tend to merge with the central, Rayleigh peak, forming in the $y < 1$ regime a broad structureless spectral feature centered at $\Omega = 0$ (dark blue and blue curves in Fig. 1a). In the $y \to 0$ limit, $S_K(\Omega)$ becomes Gaussian,

$$\lim_{y \to 0} S_K(\Omega) = \pi^{-1/2}\Omega_0^{-1} \exp(-\Omega^2/\Omega_0^2), \tag{94}$$

with $\Omega_0 = Kv_0$ as a universal property of spontaneous density fluctuations, identified already in early kinetic-theory studies of density correlations in fluids [85].

Presented in Figs. 1c and 1d are the stimulated scattering gain profiles $g_s(\Omega) = g_s(\mathbf{K}, \Omega; \mathbf{e}_p, \mathbf{e}_s)$ calculated via (80) – (85). These gain profiles are indistinguishable from the $g_s(\Omega)$ spectra computed by using the fluctuation – dissipation relation in the form of Eq. (60) with the YN solution for $S_{\mathbf{K}}(\Omega)$ as shown in Figs. 1a and 1b. The $g_s(\Omega)$ spectra in Figs. 1c and 1d are antisymmetric with respect to $\Omega = 0$, as it is required by causality, expressed by the FDT-imposed $\Omega$ factor in Eq. (60). The gain at the Stokes frequency thus becomes the loss of equal magnitude at the anti-Stokes frequency, $g_s(\mathbf{K}, \Omega; \mathbf{e}_p, \mathbf{e}_s) = -g_s(\mathbf{K}, -\Omega; \mathbf{e}_p, \mathbf{e}_s)$. This property of $g_s(\Omega)$, which is fully consistent with the available measurements [28 – 30, 109], shows that stimulated scattering from laser-driven density gratings provides a highly sensitive probe for time-resolved spectroscopic studies of gas flows and fluid dynamics, which could complement the existing toolbox of advanced spectroscopic and imaging techniques [114 – 118].

For large $y$, the $g_s(\Omega)$ spectra display well-resolved sidebands (Fig. 1d)) centered around the frequency $\Omega_B = Kv_s$, where $v_s$ is the speed of the acoustic wave. These peaks are readily identified as imprints of the signature Brillouin sidebands in spontaneous scattering (cf. Figs. 1b and 1d). The lineshapes of the Brillouin sidebands in this regime of stimulated scattering are Lorentzian, in agreement with Eq. (93). Their linewidths $\Gamma_B$ are adequately explained and accurately described in terms of the inverse phonon lifetime $\tau_B$, $\Gamma_B = \tau_B^{-1}$, as defined by the shear and dilation viscosity properties of the medium [3]. Because the Brillouin shift $\Omega_B$ and the linewidth $\Gamma_B$ in SBS are determined by material constants, SBS provides a powerful resource for the spectroscopy of a vast variety of physical, chemical, and biological systems [2, 3, 11, 119 – 129], as well as for highly sensitive bioimaging [19 – 23, 129].

Unlike the Brillouin sidebands, which accurately translate, in the large-$y$ regime, from the spectra $S_K(\Omega)$ of spontaneous density fluctuations (Fig. 1b) to SBS gain spectra (Fig. 1d), the central part of the Rayleigh peak, which is prominent in the spectra of spontaneous fluctuations, reaching its maximum at $\Omega = 0$ (Fig. 1a), is totally suppressed in the spectra of stimulated scattering (Fig. 1d). The suppression of the $\Omega = 0$ - centered part of $g_s(\Omega)$ spectra is a manifestation of the causality-driven antisymmetry of $g_s(\Omega)$ profiles, as expressed by the $\Omega$ factor in Eq. (60). The wings of the Rayleigh peak are, however, clearly visible in high-$y$ $g_s(\Omega)$ gain spectra (Fig. 1d), in agreement with and in support of the available experimental measurements [84, 98], as well as the established methodology and practice of Rayleigh-wing spectroscopy [2, 11, 12, 118].

For $y < 1$, the spacing $l_g = 2\pi/K$ of the density grating needed to satisfy momentum conservation $\mathbf{k}_p = \mathbf{k}_s + \mathbf{K}$ is smaller, $l_g < l_0$, than the mean free path $l_0$ of gas particles. Generation of sound waves with such $K$ is not possible. The Stokes wave can still be amplified in this regime via stimulated scattering off the laser-induced grating of the density of ballistic, free-streaming gas particles. As an important property of stimulated scattering in this regime, in the $y \to 0$ limit, where the spectrum of spontaneous density fluctuations becomes Gaussian (the $y = 0$ curve in Fig. 1a), the gain spectrum of stimulated scattering $g_s(\Omega)$ converges to

$$\lim_{y \to 0} g_s(\Omega) = \kappa_0(\Omega/\Omega_0^2) \exp(-\Omega^2/\Omega_0^2) = g_0(\Omega), \tag{95}$$

with $\Omega$-independent coefficient $\kappa_0$ (the $y = 0$ curve in Fig. 1c). This property of stimulated scattering has been observed in earlier studies [84, 98], but has yet to receive a clear physical explanation. Such an explanation will be provided in the next section below.

We see that the solution for the gain $g_s(\mathbf{K}, \Omega; \mathbf{e}_p, \mathbf{e}_s)$ found in Eq. (60) not only accurately recovers all the signature properties of sound-wave-mediated Stokes amplification in the large-$y$, hydrodynamic regime of stimulated scattering (Fig. 1c), but also provides a continuous crossover from the large-$y$, hydrodynamic regime to the $y < 1$, kinetic regime of stimulated scattering (Fig. 1d). Moreover, this solution offers a clear, physically intuitive explanation for distinctly different properties of stimulated scattering by density gratings in the kinetic regime, thus laying out a powerful framework for the analysis of stimulated scattering in moderate- and low-pressure gases.

## 8. Stimulated scattering via free-streaming kinetics: insights from fluctuation – dissipation relations

To gain insights into how the $g_0(\Omega)$ gain profile [Eq. (95)] emerges in the $y \ll 1$ regime of stimulated light scattering, we observe that $y \to 0$ is achieved when $\nu_c \to 0$ and examine the regime of a vanishingly low collision frequency $\nu_c$, when the collision term in the Boltzmann equation can be dropped altogether. Eq. (9) in this limit reduces to the Vlasov equation,

$$\left(\frac{\partial}{\partial t} + \mathbf{v} \cdot \boldsymbol{\nabla}\right) f + \frac{\mathbf{F}}{m} \cdot \boldsymbol{\nabla}_{\mathbf{v}} f = 0. \tag{96}$$

We search for the solution to this equation via a perturbative expansion of $f(\mathbf{r}, \mathbf{v}, t)$ as laid out in Eq. (10) with $f_0(\mathbf{r}, \mathbf{v}, t)$ taken in the form of Maxwellian distribution [Eq. (12)].

Decomposing $\mathbf{v}$ into a longitudinal and transverse velocity components, $\mathbf{v} = \mathbf{v}_\parallel + \mathbf{v}_\perp$, with $\mathbf{v}_\parallel$ chosen along $\mathbf{K}$, $\mathbf{v}_\parallel = v_\parallel \mathbf{K}/K$ (see Section 6), we express this distribution as

$$f_0(\mathbf{r},\mathbf{v},t) = f_0(\mathbf{v}) = \frac{1}{\pi^{3/2}v_0^3}\exp\left(-\frac{v^2}{v_0^2}\right) = \frac{1}{\pi^{3/2}v_0^3}\exp\left[-\frac{v_\parallel^2+v_\perp^2}{v_0^2}\right]. \quad (97)$$

Substituting Eqs. (21) – (23) into Eq. (11) with $\mathbf{F} = 0$, we find in the first order in small perturbations,

$$-\Omega\bar{f}_1 + Kv_\parallel\bar{f}_1 + \frac{\rho_0}{m}F_\Omega\frac{d\bar{f}_0}{dv_\parallel} = 0, \quad (98)$$

with $F_\Omega$ as defined in Eq. (20) and with $\bar{f}_0(v_\parallel) = \int f_0(\mathbf{v})d^2v_\perp = \frac{1}{\sqrt{\pi}v_0}\exp\left(-\frac{v_\parallel^2}{v_0^2}\right)$ and $\bar{f}_{1,K,\Omega}(v_\parallel) = \int d^2v_\perp \bar{f}_{1,K,\Omega}(\mathbf{v})$, such that $\int \bar{f}_0(v_\parallel)\,dv_\parallel = 1$.

Solving Eq. (98) for $\bar{f}_1 = \bar{f}_{1,K,\Omega}(v_\parallel)$ yields

$$\bar{f}_{1,K,\Omega}(v_\parallel) = \frac{\rho_0 F_\Omega}{m(\Omega - Kv_\parallel)}\frac{d\bar{f}_0}{dv_\parallel}. \quad (99)$$

Electrostriction-induced density change is now found as $\delta\rho_{\mathbf{K},\Omega} = \int \bar{f}_{1,K,\Omega}(v_\parallel)\,dv_\parallel$, leading to

$$\delta\rho_{\mathbf{K},\Omega} = \int \frac{K\mathcal{U}_0}{\Omega - Kv_\parallel}\frac{d\bar{f}_0}{dv_\parallel}dv_\parallel. \quad (100)$$

Eq. (100) can be expressed in the form of Eq. (48),

$$\delta\rho_{\mathbf{K},\Omega} = \rho_0 F_\Omega\, \mathcal{Q}(K,\Omega), \quad (101)$$

with

$$\mathcal{Q}(K,\Omega) = \frac{1}{m}\int_{-\infty}^{\infty}\frac{d\bar{f}_0/dv_\parallel}{\Omega - Kv_\parallel}dv_\parallel \quad (102)$$

recognized as the spectrum of the response function [Eqs. (46), (47)].

Expanding the denominator in the integrand in Eq. (102) via the Sokhotski–Plemelj theorem,

$$\lim_{\epsilon\to 0^+}\frac{1}{\Omega - Kv_\parallel + i\epsilon} = \mathcal{P}\left(\frac{1}{\Omega - Kv_\parallel}\right) - i\pi\delta(\Omega - Kv_\parallel), \quad (103)$$

with $\mathcal{P}(\cdot)$ standing for the Cauchy principal-value integral, we find for the imaginary part of the cubic susceptibility $\chi^{(3)}$ [Eq. (58)]:

$$\mathrm{Im}\left[\chi^{(3)}\right] = -\frac{1}{8\pi^{3/2}}\frac{{\gamma_e}^2}{\rho_0 v_0^2}\theta\exp(-\theta^2), \quad (104)$$

with $\theta = \frac{\Omega}{Kv_0}$.

The stimulated scattering gain can now be found via Eqs. (59) and (60), yielding

$$g_s = \frac{\sqrt{\pi}\,\omega_s}{n_p n_s c^2}\frac{\gamma_e{}^2}{\rho_0 v_0^2}\theta \exp(-\theta^2), \tag{105}$$

or, in terms of polarizability $\alpha$,

$$g_s = \frac{8\pi^{5/2}\omega_s}{n_p n_s c^2}\frac{\alpha^2 \mathcal{N}}{k_B T}\theta \exp(-\theta^2). \tag{106}$$

Eqs. (105) and (106) offer important insights into the emergence of the $g_0(\Omega)$ gain profile [Eq. (95)] as the universal $y \to 0$ limit for the stimulated scattering gain spectrum. Indeed, the $g_0(\Omega)$ spectral profile, readily recognizable as the Rayleigh distribution [Eq. (95)], is seen to be an expression of the $(\Omega - Kv_{\parallel})^{-1}$ singularity in the $\overline{f}_{1,K,\Omega}(v_{\parallel})$ correction to the one-particle distribution function [Eq. (99)] as dictated by the collisionless, $\nu_c \to 0$ limit of the Boltzmann equation. Such a singularity is a clear signature of the $\Omega = Kv_{\parallel}$ resonance in the response of the system to the electrostriction-driven external force, captured by the solution for $\overline{f}_{1,K,\Omega}(v_{\parallel})$ found in Eq. (99).

The prominence of the $\Omega = Kv_{\parallel}$ resonance in low-$y$ stimulated scattering is representative of the physical inner workings behind the kinetic regime of stimulated light scattering. While in the large-$y$, hydrodynamic regime, stimulated scattering is enabled by the interaction of the laser field with acoustic waves, in the $y < 1$ regime, the spacing $l_g = 2\pi/K$ of the density grating needed to phase-match stimulated Stokes amplification is smaller than the mean free path of gas particles. Acoustic waves are thus excluded as mediators of stimulated scattering. However, an exponential buildup of the Stokes field in this regime is still possible, enabled by stimulated scattering off laser-induced gratings of the density of ballistic, free-streaming gas particles. In such gratings, gas particles whose velocity $\mathbf{v}$ satisfies $\mathbf{K}\cdot\mathbf{v} = \Omega$ are in phase with the driver force $\mathbf{F}(\mathbf{r}, \mathrm{t})|_{\mathbf{K},\Omega}$ as defined by Eq. (20). Eq. (98) is instructive in isolating the $-\Omega\overline{f}_1$ and $Kv_{\parallel}\overline{f}_1$ terms via differentiation in time and velocity, as prescribed by the collisionless Boltzmann equation (96). The sum of these terms then enters into the solution for $\overline{f}_{1,K,\Omega}(v_{\parallel})$ as a resonant denominator $(\Omega - Kv_{\parallel})^{-1}$ ([Eq. (99)].

We see that, physically, the emergence of the $g_0(\Omega)$ profile as a universal, model-insensitive $\nu_c \to 0$ limit for $g_s(\Omega)$ (the $y = 0$ curve in Fig. 1c) indicates that all disturbances and fluctuations dissipate in this regime via a flow of uncorrelated free-streaming particles, all propagating with their individual thermal velocities. The spontaneous-scattering counterpart of the $g_0(\Omega)$ gain profile is a Gaussian profile of $S(\Omega)$, emerging as a universal, and physically intuitive, $\nu_c \to 0$ limit for the spectrum of spontaneous density fluctuations (the $y = 0$ curve in Fig. 1a), which, in the absence of collisions, dissipate as a part of uncorrelated kinetics of free-streaming particles.

When achieved via this procedure, the solution for $\delta\rho_{\mathbf{K},\Omega}$ [Eqs. (100) – (102)] offers a physically clear view of how the fluctuation – dissipation connection sets in in the collisionless regime. In the absence of collisions, kinetics of individual gas particles are totally uncorrelated. Any disturbance in gas density, however, induced by an external impulsive force or arising spontaneously, will still gradually fade away, dissipating via an uncorrelated ballistic motion of gas particles, all moving with their individual velocities. At equilibrium, the distribution of these velocities is Maxwellian, leading to the Gaussian spectrum as a universal, model-nonspecific $\nu_c \to 0$ limit for uncorrelated density fluctuations [Eq. (94)]. Because Stokes gain (or anti-Stokes loss) in stimulated light scattering by laser-induced density gratings is the dissipation counterpart to spontaneous density fluctuations, it connects to the spectrum of spontaneous density fluctuations via Eqs. (100) – (106), in full agreement with the fluctuation – dissipation theorem.

**9. Quantitative treatment**

The properties of stimulated gain spectra $g_s(\Omega)$ in the kinetic, $y < 1$ regime are thus distinctly different from the properties of SBS gain spectra in the $y \gg 1$ regime. In the $y < 1$ regime, Eq. (9) is no longer reducible to the Navier – Stokes equation. The properties of the medium are therefore no longer understood in terms of hydrodynamics of continuous fluid. Specifically, the velocity $\mathbf{v}$ no longer translates into the velocity of acoustic modes but is understood instead in a more general sense of the kinetic framework of Eqs. (9) – (14) as a random velocity of gas particles whose distribution function is found by solving these equations. Unlike the standard hydrodynamic solution for the SBS gain profile [2 – 5], in which the gain spectrum peaks at $\Omega_B = Kv_s$ and the gain bandwidth $\Gamma_B$ is determined by the phonon lifetime $\tau_B$, the position $\Omega_s$ of the maximum in the $y < 1$ gain spectrum and the bandwidth of this spectrum are decoupled, as Eqs. (60) and (105) show, from any parameters that would characterize the gas as a continuous medium, including the viscosity and the speed of any acoustic wave.

Indeed, searching for the maximum $\Omega_s$ of the $g_s(\Omega)$ gain spectrum in the $y \to 0$ limit, we find $\Omega_s = Kv_0/\sqrt{2}$. The full width of the $g_s(\Omega)$ gain band at the $1/e$ level in the $y \to 0$ limit, on the other hand, is $\Delta\Omega_{1/e} \approx 1.34Kv_0$. We see that, rather than connecting to the speed of sound and the phonon lifetime, the frequency of the maximum in the $y < 1$ $g_s(\Omega)$ gain spectrum and the bandwidth of this spectrum are determined by distinctly different factors, viz., the product $Kv_0$, as an expression of a distinctly different physics behind stimulated scattering in the $y < 1$ regime. Specifically, while the linewidth $\Gamma_B$ emerging in the $y \gg 1$, hydrodynamic regime of stimulated Brillouin scattering is determined by the phonon lifetime $\tau_B$, the gain bandwidth $\Delta\Omega_{1/e}$ in the $y \to 0$ regime of stimulated scattering by density gratings connects, via $\Delta\Omega_{1/e} \approx 1.34Kv_0$ with $Kv_0 = 2\pi/\tau_s$, to the ballistic time $\tau_s = l_g/v_0$, i.e., the time required for a free-streaming gas particle with a thermal velocity $v_0$ to travel a distance $l_g = 2\pi/K$ equal to the spacing of the density grating needed to phase-match stimulated light amplification. The time $\tau_s$ is, by its definition, a measure of the dephasing time of free-streaming ballistic gas particles.

Typical estimates for the room-temperature thermal velocity $v_0$ in gases commonly used in stimulated scattering experiments range from 180 m/s for $SF_6$, 190 m/s for Xe, and 240 m/s for Kr to 420 m/s for $N_2$ and 500 m/s for Ne, reaching 1100 m/s for He. The gain bandwidth achieved in backward stimulated scattering of $\lambda_p \approx 266$ nm laser radiation is thus estimated as $\Delta\Omega_{1/e}/(2\pi) \approx 1.8$ GHz for $SF_6$, ≈ 2.4 GHz for Kr, and ≈ 4.2 GHz for $N_2$. The frequency of the maximum in the gain spectrum is $\Omega_s/(2\pi) \approx 0.93$ GHz for $SF_6$, ≈ 1.3 GHz for Kr, and ≈ 2.2 GHz for $N_2$. These estimates agree very well with the available measurements [28 – 30, 66 – 69, 84, 98].

The solution for the stimulated gain $g_s(\Omega)$ found in Eqs. (105) and (106) provides important insights into the key physical factors that define the absolute value of $g_s$ and its behavior as a function of the gas pressure. To understand these factors, we first search for the maximum of $g_s(\Omega)$ over the gain spectrum by differentiating $\theta \exp(-\theta^2)$ in Eqs. (105) and (106) in $\theta$ and equating this derivative to zero. This search returns $\theta_m = 1/\sqrt{2}$. $g_s(\Omega)$ thus reaches its maximum at $\Omega_m = Kv_0/\sqrt{2}$, where $(\Omega_m/Kv_0)\exp[-(\Omega_m/Kv_0)^2] = 1/\sqrt{2e} \approx 0.43$, leading, via Eqs. (105) and (106) with $n_p \approx n_s \approx 1$, to

$$g_m = g_s(\Omega_m) \approx 0.43\frac{\sqrt{\pi}\omega_s}{c^2}\frac{{\gamma_e}^2}{\rho_0 v_0^2} = 0.43\frac{8\pi^{5/2}\omega_s}{c^2}\frac{\alpha^2\mathcal{N}}{k_B T}. \quad (107)$$

As a meaningful reference, for backward stimulated scattering of laser radiation with $\lambda \approx$ 266 nm in room-temperature Ar at 1 atm, $\alpha \approx 1.8$ Å$^3$, Eq. (107) gives $g_m \approx 10$ cm/TW. This estimate agrees very well with spectrally resolved studies of backward stimulated scattering of 266-nm laser radiation in room-temperature Ar at 1 atm, yielding the gain value of ≈ 10 cm/TW at the maximum of the gain spectrum.

When expressed in the form of Eq. (106), the small-$y$ solution for $g_s$ is especially transparent in revealing the behavior of $g_s$ as a function of the gas pressure. Indeed, with the gas temperature and the frequency of the laser driver fixed, $g_s$ in Eq. (106) scales as $\alpha^2\mathcal{N}$ with polarizability $\alpha$ and the number density $\mathcal{N}$. This linear dependence on $\mathcal{N}$, observed in the small-$y$ regime, is in striking contrast with the behavior of the large-$y$ SBS gain, found via a standard, hydrodynamic treatment of $y \gg 1$ SBS [2, 3, 5, 11], $g_s \propto \frac{\gamma_e{}^2}{\rho_0\Gamma_B}$. Expressing the Brillouin linewidth as $\Gamma_B \approx (2\eta_s + \eta_d)K^2/\rho_0$ in terms of the shear and dilation viscosity coefficients $\eta_s$ and $\eta_d$ [2, 3, 5, 11], we can recast the $y \gg 1$ solution for the SBS gain as $g_s \propto \alpha^2\mathcal{N}^2$. Unlike the small-$y$, kinetic $g_s$, which scales linearly with $\mathcal{N}$, the $y \gg 1$ , hydrodynamic SBS gain is a quadratic function of $\mathcal{N}$. This finding is also fully consistent with experimental studies of the gas-pressure dependence of stimulated Stokes amplification, performed as a part of SBS research since the early days of nonlinear optics (e.g., Refs. 6, 7, 119). As a useful reference for the hydrodynamic regime, the Stokes gain as high as ≈ 25 cm/GW is attainable for 250-nm laser radiation in $SF_6$ at 15.5 atm [67, 68], enabling highly efficient pulse compression and amplification, as well as beam combining in high-energy, high-intensity laser beamlines.

## 10. Conclusion

To summarize, when integrated into a suitable kinetic framework, fluctuation–dissipation relations provide a powerful resource for the analysis of stimulated light scattering from laser-driven density gratings, allowing the stimulated gain/loss spectrum to be found directly from the spectrum of spontaneous density fluctuations without the need to solve kinetic equations with an external-field term. At the conceptual level, the fluctuation–dissipation treatment of stimulated light scattering establishes a connection between spontaneous fluctuations of material properties and stimulated optical gain/loss in nonlinear light – matter interactions. In density-modulation-mediated scattering, this connection is an imprint of the inner physical workings behind light–matter interaction: the laser-driven beat wave induces a density perturbation, giving rise to an electrostrictive force, which couples the optical fields to material density waves, thus opening a channel for Stokes-field amplification. Although the material response behind this process is cubic in the laser field [Eq. (6)], it is linear in the electrostrictive force [Eq. (5)], assuming its third-order nonlinearity via the quadratic field dependence of ponderomotively driven electrostriction [Eqs. (3), (4)]. Thus, even though stimulated light scattering is rightfully viewed as one of archetypal nonlinear-optical processes [1 – 3, 5 – 12], extension of fluctuation – dissipation relations to this class of light – matter interactions does not require a nonlinear FDT generalization, nor does it lead to any form of nonlinear FDT, but can operate, as our study shows, with standard FDT formulations applied to a material response that enters linearly into a nonlinear optical interaction.

As an important practical aspect, the FDT treatment presented in this study provides a framework whereby spontaneous-scattering data can be meaningfully fed into the analysis of stimulated and coherent scattering, suggesting means to understand and predict the key properties of stimulated gain and loss spectra as functions of light – matter interaction parameters. While in molecular spectroscopy, such connection between nonlinear and spontaneous Raman scattering is well appreciated [2, 3, 11 – 13], in strong-field laser – plasma interactions, including laser-fusion physics, it is a subject of ongoing intense research [130 – 132].

Because the FDT, in its canonical form, is applicable to systems at equilibrium, applications of FDT-based methodologies are limited largely to systems fluctuating around and/or being driven from their equilibria. Furthermore, while the FDT establishes sometimes nontrivial connections between spontaneous fluctuations and the linear response of a driven system, it is in no way intended as a substitute for the physical model of the material response, such as the model of ponderomotively driven electrostriction in stimulated light scattering examined in this study. Finally, falling beyond the scope of FDT-only treatment are effects related to the spatiotemporal evolution of optical fields and coupled material waves, including ultrafast dynamics of ultrashort laser pulses, material-response buildup, as well as effects of boundaries, inhomogeneities, particle trapping, and wave breaking, playing a prominent role in laser – plasma interactions.

**Acknowledgments**
This research was supported in part by the DOE Office of Science through the DOE Inertial Fusion Energy Science and Technology Accelerator Research (IFE-STAR) program (grant # DE-SC0024882).

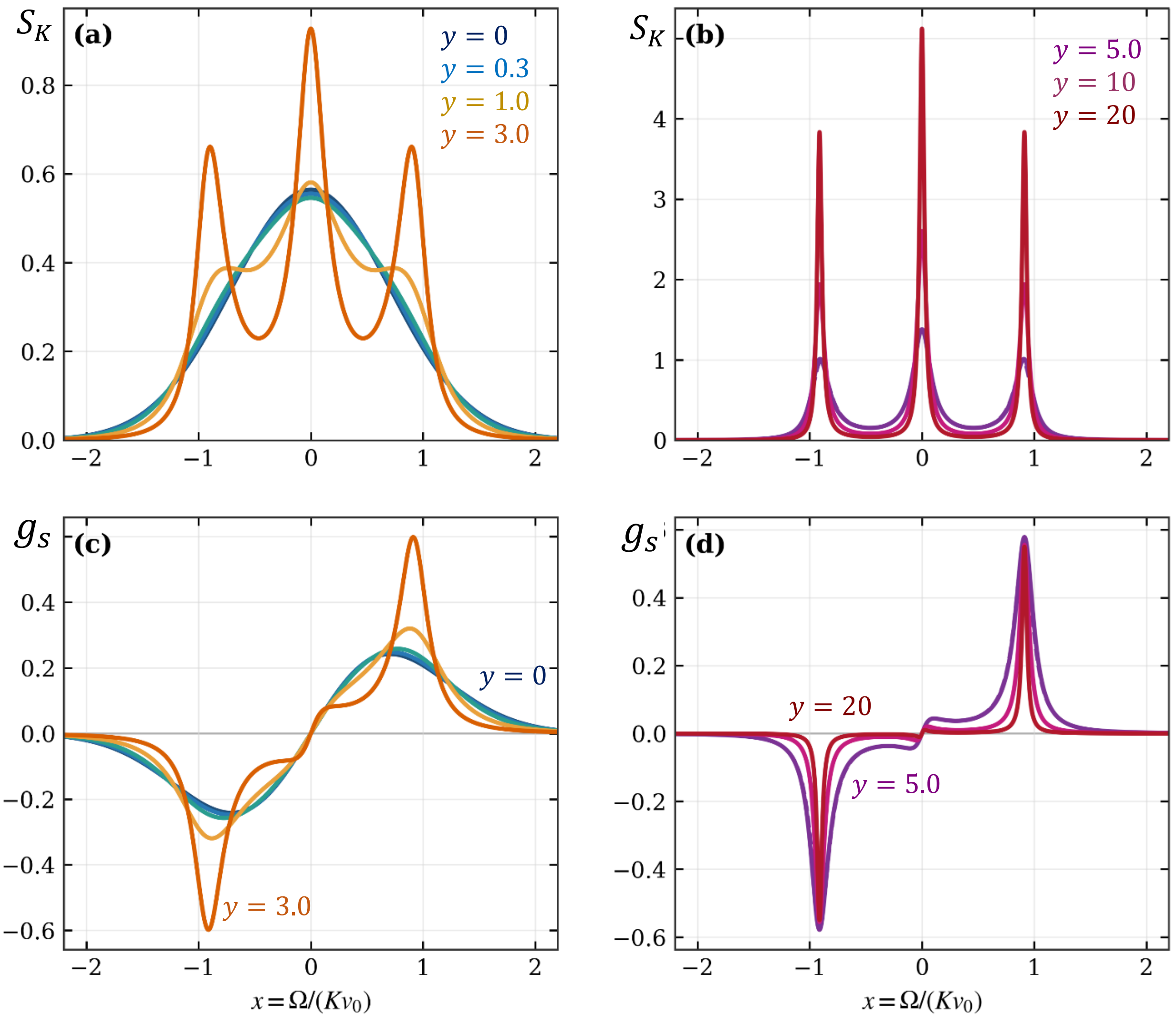

Fig. 1. (a, b) The spectral density $S_K$ of the equilibrium correlation function of spontaneous density fluctuations calculated using the Yip – Nelkin solution [85, 88] to the Boltzmann equation with the linearized BGK collision term [Eqs. (11) – (14)] without an external force, $\mathbf{F} = 0$, and (c, d) stimulated scattering gain $g_s$ calculated using Eqs. (80) – (85). The spectral densities $S$ and $g_s$ are plotted as functions of $x = \frac{\Omega}{Kv_0}$ for (a, c) $y = \nu_c/(Kv_0) = 0$ (dark blue), 0.3 (blue), 1 (gold), and 3 (orange) and (b, d) $y = 5$ (purple), 10 (magenta), and 20 (dark red).